# Mobility in N-Doped Wurtzite III-Nitrides

*C.G. Rodrigues*[a]*, *Valder N. Freire*[b], *Áurea R. Vasconcellos*[c], *Roberto Luzzi*[c]

[a]*Núcleo de Pesquisa em Física, Departamento de Física,*
*Universidade Católica de Goiás, C.P. 86, 74605-010 Goiânia - GO, Brazil*
[b]*Departamento de Física, Universidade Federal do Ceará,*
*C.P. 6030, Campus do Pici, 60455-760 Fortaleza - CE, Brazil*
[c]*Instituto de Física Gleb Wataghin, Universidade Estadual de Campinas,*
*C.P. 6165, 13083-970 Campinas - SP, Brazil*
`cloves@pucgoias.edu.br`

A study of the mobility of *n*-doped wurtzite III-Nitrides is reported. We have determined the nonequilibrium thermodynamic state of the III-Nitrides systems – driven far away from equilibrium by a strong electric field – in the steady state, which follows after a very fast transient. The dependence of the mobility (which depends on the nonequilibrium thermodynamic state of the sample) on the electric field strength is derived, which decreases with the strength of electric field. We analyzed the contributions to the mobility arising out of the different channels of electron scattering, namely, the polar optic, deformation, piezoelectric, interactions with the phonons, and with impurities. The case of n-InN, n-GaN, and n-AlN have been analyzed: as expected the main contribution comes from the polar-optic interactions in these strongly polar semiconductors. The other interactions are in decreasing order, the deformation acoustic, the piezoelectric, and the one due to impurities.

## 1. Introduction

There exists nowadays a large interest in the study of large-gap semiconductors – like the III-Nitrides – because of their use in devices, like diodes and lasers, emitting in the blue and near-ultraviolet region[1]. Optical and transport properties are being investigated, and we have concentrated our attention on these properties when the doped semiconductor is in the presence of moderated to high electric fields. In particular we report here a study of the mobility of n-doped III-Nitrides in such conditions. For that purpose we have determined the nonequilibrium thermodynamic state of the system – driven far-away from equilibrium by the electric field – in the steady state, which follows after a very rapid (picosecond scale) transient[2,3]. The drift velocity (and hence the current) is derived – which of course depends on the nonequilibrium macroscopic (thermodynamic) state of the sample. The dependence of the mobility with the electric field strength is obtained, which decreases with increasing strength of the electric field. We have also analyzed the contributions to the mobility arising out of the different channels of electron scattering, namely, the polar optic, deformation, piezoelectric, interactions with the phonons and with impurities. The case of n-InN, n-GaN, and n-AlN have been analyzed: as expected at the room temperature, the main contribution comes from the polar-optic interactions in these strongly polar semiconductors. The other interactions are in decreasing order, the deformation acoustic, the piezoelectric, and the one due to impurities. Only in the case of n-AlN, the deformation acoustic contribution is roughly 10% of the polar optic; in the other two is of the order of 1 to 2%.

## 2. Mobility of Electrons

Consider the case of *n*-doped direct-gap polar semiconductors, in contact and in equilibrium with a reservoir at temperature $T_0$ (300 K in the numerical calculations to be presented later on). An electric field $F$ is applied, in, say, *x* direction. It drives the system out of equilibrium, and the time-dependent (due to the relaxation processes that unfold in it) macroscopic state is described in terms of a statistical thermodynamics for irreversible systems, namely Informational Statistical Thermodynamics[4], which is based on the so-called Predictive Statistical Mechanics[5]. In the present case, the nonequilibrium thermodynamic state is characterized by the macrovariables carriers' energy, $E_e(t)$, carriers' linear momentum, $P_e(t)$, along the $x$-axis, and the energies of the longitudinal optical and acoustic phonons, $E_{LO}(t)$ and $E_{AC}(t)$ respectively[6] (all are given per unit volume). The TO phonons have been ignored once it is negligible the deformation potential interaction with electrons in the conduction band.

The equations of evolution for these basic macrovariables are derived in the nonlinear quantum kinetic theory described in Ref. 7. They are

$$\frac{d}{dt}E_e(t) = -\frac{eF}{m_e^*}P_e(t) + J_E^{(2)}(t) , \qquad (1)$$

$$\frac{d}{dt}P_e(t) = -nVeF + J_{P_e}^{(2)}(t) + J_{P_e,imp}^{(2)}(t) , \qquad (2)$$

$$\frac{d}{dt}E_{LO}(t) = -J_{E_{LO}}^{(2)}(t) - J_{LO,AN}^{(2)}(t) , \qquad (3)$$

$$\frac{d}{dt}E_{AC}(t) = -J_{E_{AC}}^{(2)}(t) + J_{LO,AN}^{(2)}(t) - J_{AC,dif}^{(2)}(t) \qquad (4)$$

(all $J^{(2)}$ being positive or null only when equilibrium is attained). $E_e$ is the electrons' energy and $P_e$ their linear momentum; $E_{LO}$ the energy of the LO phonons, which strongly interact with the carriers via Fröhlich potential in these strong-polar semiconductors; $E_{AC}$ is the energy of the acoustic phonons, which play a role of a thermal bath; and $F$ stands for the constant electric field in the $x$-direction.

Let us analyze these equations term by term. In Eq.1 the first term on the right accounts for the rate of energy transferred from the electric field to the carriers, and the second term accounts for the transfer of the excess energy of the carriers – received in the first term – to the phonons. In Eq. 2 the first term on the right is the driving force generated by the presence of the electric field. The second term is the rate of momentum transfer due to interaction with the phonons, and the last one is a result of scattering by impurities. In Eq. 3 and Eq. 4 the first term on the right describes the rate of change of the energy of the phonons due to interaction with the electrons. More precisely they account for the gain of the energy transferred to then from the hot carriers and then the sum of contributions $J_{E_{LO}}^{(2)}$ Jand $J_{E_{AC}}^{(2)}$ is equal to the last in Eq. 1, with change of sign. The second term in Eq. 3 accounts for the rate of transfer of energy from the optical phonons to the acoustic ones via anharmonic interaction. The contribution $J_{LO,AN}^{(2)}$ is the same but with different sign in Eq. 3 and 4. Finally, the diffusion of heat from the AC phonons to the reservoir is account for in the last term in Eq. 4. The detailed expressions for the collision operators are given in Ref. 6.

We notice that the linear momentum density can by related to the drift velocity $v_e(t)$ (along the $x$-axis) by the relation

$$P_e(t) = Nm_e^* v_e(t) \qquad (5)$$

where $m_e^*$ is the effective mass of the electron, related to the current $I_e$ by the expression

$$I_e(t) = -nev_e(t) \qquad (6)$$

which flows in the direction of the electric field. Moreover we define a momentum relaxation time, associated to the scattering of carriers by phonons, given by

$$\tau_{P_e}(t) \equiv -n\frac{m_e^* v_e(t)}{J_{P_e}^{(2)}(t) + J_{P_e,imp}^{(2)}(t)} \qquad (7)$$

where, once we take into account that the collision operator $J_{P_e}^{(2)}$ is composed of the three contributions consisting of scattering by: optical (or Fröhlich) interaction with LO phonons, deformation potential with LO and AC phonons, and through the piezoelectric potential with AC phonons. We have then a Mathiessen-like rule of the form

$$\frac{1}{\tau_{P_e}} = \frac{1}{\tau_{PO}} + \frac{1}{\tau_{AD}} + \frac{1}{\tau_{PZ}} + \frac{1}{\tau_{imp}} \qquad (8)$$

In the Eq. 8, *PO* refers to polar-optic interaction, *AD* to acoustic deformation potential, *PZ* to acoustic piezoelectric potential, and *imp* to the effect of impurities. We have introduced the individual relaxation times

$$\tau_i = \frac{nm_e^* v_e}{J_i^{(2)}} \qquad (9)$$

where $J_i^{(2)}$ is the contribution to of Eq. 4 for $i = PO$, *AD*, *PZ*, and, $i = imp$ for $J_{imp}^{(2)}$. Consequently, the mobility in the steady state, namely,

$$M_e = \frac{v_e}{F} = \frac{e}{m_e^*}\tau_{P_e} \quad (10)$$

(once, according to Eq. 4, in the steady state, $J_{P_e}^{(2)} + J_{P_e,imp}^{(2)} = neF$ ) becomes a composition of four contributions, according to the rule

$$\frac{1}{M_e} = \frac{1}{M_{PO}} + \frac{1}{M_{AD}} + \frac{1}{M_{PZ}} + \frac{1}{M_{imp}} \quad (11)$$

## 3. Results

Numerical results are shown in Figs. 1-4. We can see that InN the highest mobility, followed by GaN and AlN. Inspection of Table 1 tells us that this is a consequence, as expected, from the fact that the effective mass of the carriers follows mainly due to the ordering $m_{InN}^* < m_{GaN}^* < m_{AlN}^*$.

For the same three III-Nitrides, Figs. 2-4 shown a comparison of the four contributions to the mobility cf. Eq. 11. Notice that is valid Mathiessen rule, and then the much smaller practically determines the value of the whole mobility. This is clearly a consequence that polar-optic (Fröhlich) interaction is strong in this compounds, and therefore produces a very short relaxation time as compared with the other scattering mechanisms (deformation acoustic and piezoelectric). Also it can be noticed that in the conditions of the calculations, the scattering by the ionized impurities is negligible when compared to the others.

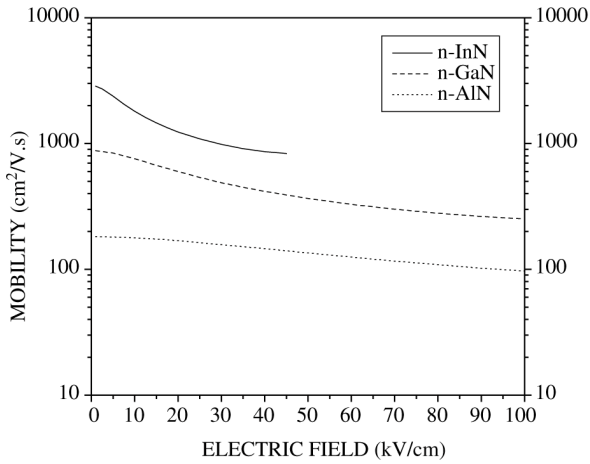

**Figure 1.** Electric field dependence of the electron mobility in wurtzite n-InN (solid line), n-GaN (dashed line), and n-AlN (doted). The lattice temperature is $T_0 = 300$ K and $n = 10^{17}$ cm$^{-3}$.

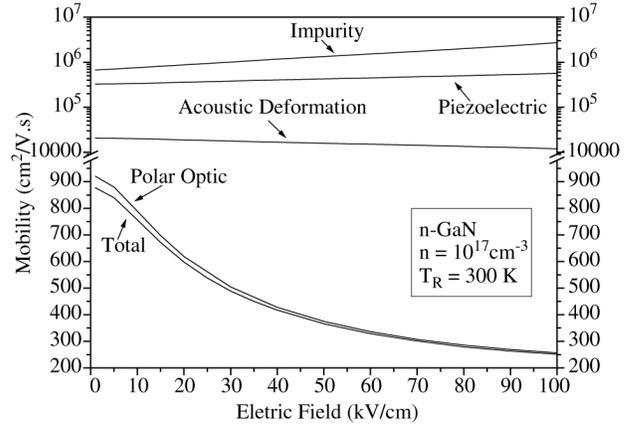

**Figure 3.** Electric field dependence of the different contributions to the electron mobility in wurtzite n-GaN. The lattice temperature is $T_0 = 300$ K and $n = 10^{17}$ cm$^{-3}$.

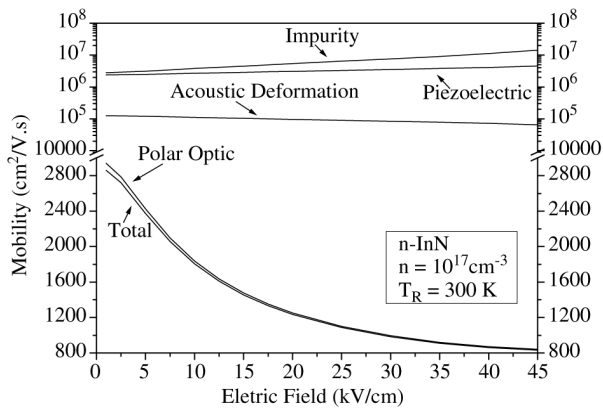

**Figure 2.** Electric field dependence of the different contributions to the electron mobility in wurtzite n-InN. The lattice temperature is $T_0 = 300$ K and $n = 10^{17}$ cm$^{-3}$.

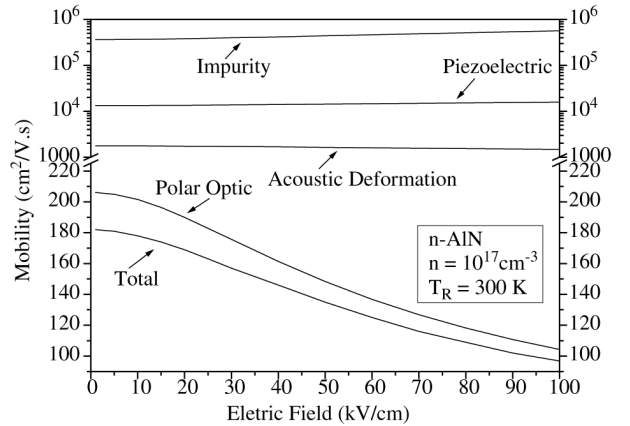

**Figure 4.** Electric field dependence of the different contributions to the electron mobility in wurtzite n-AlN. The lattice temperature is $T_0 = 300$ K and $n = 10^{17}$ cm$^{-3}$.

**Table 1.** Parameters of wurtzite AlN, InN, and GaN used in the numerical solution of the transport equations.

| Parameter | AlN | InN | GaN |
|---|---|---|---|
| electron effective mass $m_e^*$ ($m_0$) | 0.35[a] | 0.11[d] | 0.19[a] |
| lattice constant a (Å) | 3.11[b] | 3.54[e] | 3.189[f] |
| lattice constant c (Å) | 4.98[b] | 5.7[e] | 5.185[f] |
| LO-phonon energy $\eta\omega_{LO}$ (meV) | 99.2[b] | 89[e] | 92[g] |
| optical dielectric constant, $\varepsilon_\infty$ | 4.77[b] | 8.4[e] | 5.35[f] |
| static dielectric constant, $\varepsilon_0$ | 8.5[b] | 15.5[e] | 9.5[f] |
| mass density (g/cm$^3$) | 3.23[b] | 6.81[e] | 6.09[h] |
| sound velocity ($\times 10^5$ cm/s) | 6.04[b] | 4.16[b] | 4.40[b] |
| acoustic deformation potential $E_{1e}$ (eV) | 9.5[b] | 7.1[e] | 8.3[i] |
| piezoelectric constant $h_{PZ}$ (C/m$^2$) | 0.92[c] | 0.375[e] | 0.375[i] |

[a]Ref. [8], [b]Ref. [9], [c]Ref. [10], [d]Ref. [11], [e]Ref. [12], [f]Ref. [13], [g]Ref. [14], [h]Ref. [15], [i]Ref. [16].

## Acknowledgement

The authors acknowledge financial support received from the São Paulo State Research Agency (FAPESP), the Brazilian National Research Council (CNPq), the Ceará State Research Funding Agency (FUNCAP), and the Brazilian Ministry of Planning through FINEP; ARV, VNF, and RL are CNPq Research Fellows. The author CGR acknowledge financial support received from the Goiás State Research Agency (SECTEC-proc. 166/003-2001) and CNPq (proc. 472354/2001-9).

## References

1. Nakamura, S.; Fasol, G. *The Blue Laser Diode* (Springer, Berlin, 1997); Mohammad, S. N. Morkoç, H., Prog. Quantum Electron., v. 20, p. 361, 1996; Akasaki, I., Amano, H., Jpn. *J. Appl. Phys.*, v. 36, p. 5393, 1997, and references therein.
2. Rodrigues, C.G.; Freire, V.N.; Costa, J.A.P.; Vasconcellos, A.R.; Luzzi, R. *Phys. Stat. Sol. (b)*, v. 216, n. 1, p. 35-39, 1999.
3. Rodrigues, C.G.; Freire, V.N.; Vasconcellos, A.R.; Luzzi, R. *Applied Physics Letters*, v. 76, n. 14, p. 1893-1895, 2000.
4. Luzzi, R.; Vasconcellos, A.R.; Ramos, J.G.; *Statistical Foundations of Irreversible Thermodynamics*, (Teubner-BertelsmannSpringer, Stuttgart, 2000); also, Rivista Nuovo Cim., v. 24, n. 3, p. 1-70, 2001.
5. Luzzi, R.; Vasconcellos, A.R.; Ramos, J.G.; *Predictive Statistical Mechanics: A Nonequilibrium Ensemble Formalism*, (Kluwer Academic, Dordrecht, in press); also a review at E-Print xxx.lanl.gov/cond-mat/9909160, 1999; Fortschr. Phys./Prog. Phys., v. 38, p. 887, 1990; ibid, v. 43, p. 265, 1995; J. Mod. Phys. B, v. 14, p. 3189, 2000.
6. Rodrigues, C.G., Ph.D. Thesis, Campinas State University (Campinas, São Paulo, Brazil, unpublished, 2001).
7. Rodrigues, C.G.; Vasconcellos, A.R.; Luzzi, R. *Transp. Theory and Stat. Phys.*, v. 29, n. 7, p. 733-757, 2000.
8. Kim, K. *et al.*, *Phys. Rev. B*, v. 56, n. 12, p. 7363, 1997.
9. Chin, V.W.L. *et al.*, *J. Appl. Phys.*, v. 75, p. 7365, 1994.
10. O'Leary, S.K. *et al.*, *Sol. Stat. Commun.*, v. 105, p. 621, 1998.
11. Yeio, Y. C. *et al.*, *J. Appl. Phys.*, v. 75, p. 1429, 1998.
12. Yim, W.M. *et al.*, *J. Appl. Phys.*, v. 44, p. 292, 1973.
13. Strite, S.; Morkoç, H. *J. Vac. Sci. Technol. B*, v. 10, p. 1237, 1992.
14. Baker, A. S.; Ilegems, M. *Phys. Rev. B*, v. 7, p. 743, 1973.
15. Hahn, H.; Juza, R.; *Z. Allg. Anorg. Chem.*, v. 244, p. 111, 1940.
16. Shur, M. *et al.*, *J. Electronic Materials*, v. 25, n. 5, p. 777, 1996.